\documentclass[manuscript]{emulateapj}
\def\nh{$N_{H}$}
\def\eg{{\it e.g.}}

\def\ergcms{\,\text{erg}\,\text{cm}$^{-2}$\,\text{s}$^{-1}$}
\def\chis{{$\chi^2_{\nu}$}}
\def\chiss{{$\chi^2$}}
\def\ergs{\,erg\,s$^{-1}$}
\def\ergcms{\,erg\,cm$^{-2}$\,s$^{-1}$ }
\def\diskbb{{\sc diskbb}}
\def\isis{{\sc ISIS}}
\newcommand{\integral}{\textsl{INTEGRAL}}

\newcommand{\chandra}{{\it \mbox{Chandra}}}
\newcommand{\swift}{{\it \mbox{Swift}}}

\newcommand{\mysou}{{\objectname{IGR~J17497$-$2821}}}

\shorttitle{Hunting the nature of \mysou~with X-ray and NIR observations}
\shortauthors{Paizis, Nowak, Chaty et al. 2007}

\begin{document}


\title{Hunting the nature of \mysou~with X-ray and NIR observations}


\author{A. Paizis\altaffilmark{1}}  
\email{ada@iasf-milano.inaf.it}
\author{M. A. Nowak\altaffilmark{2}}
\author{S. Chaty\altaffilmark{3}}
\author{J. Rodriguez\altaffilmark{3}}
\author{T.~J.-L. Courvoisier\altaffilmark{4,}\altaffilmark{5}}
\author{M. Del Santo\altaffilmark{6}}
\author{K. Ebisawa\altaffilmark{7}}
\author{R. Farinelli\altaffilmark{8}}
\author{P. Ubertini\altaffilmark{6}}
\author{J. Wilms\altaffilmark{9}}
\altaffiltext{1}{IASF Milano - INAF, Via Bassini 15, 20133 Milano, Italy}
\altaffiltext{2}{Center for Space Research, MIT, Cambridge, MA, USA, mnowak@space.mit.edu}
\altaffiltext{3}{AIM - Astrophysique Interactions Multi-\'echelles (UMR 7158 CEA/CNRS/Universit\'e Paris 7 Denis Diderot)
CEA Saclay, DSM/DAPNIA/Service d'Astrophysique, FR-91 191 Gif-sur-Yvette Cedex, France, chaty@cea.fr, rodrigue@discovery.saclay.cea.fr}
\altaffiltext{4}{\textit{INTEGRAL} Science Data Centre, Chemin d'Ecogia 16, 1290 Versoix, Switzerland,  Thierry.Courvoisier@obs.unige.ch}
\altaffiltext{5}{Observatoire de Gen\`eve, 51 chemin des Mailletes, 1290 Sauverny, Switzerland}
\altaffiltext{6}{IASF Roma - INAF, Via del Fosso del Cavaliere 100, 00133 Roma, Italy,  melania.delsanto@iasf-roma.inaf.it, pietro.ubertini@iasf-roma.inaf.it}
\altaffiltext{7}{ISAS 3-1-1 Yoshinodai, Sagamihara, Kanagawa 229-8510, Japan, ebisawa@astro.isas.jaxa.jp}
\altaffiltext{8}{Dipartimento di Fisica, Universit\`{a} di Ferrara, Via Saragat 1, I--44100 Ferrara, Italy, farinel@fe.infn.it }
\altaffiltext{9}{Dr. Karl Remeis-Sternwarte, Astronomisches Institut, Universit\"at Erlangen-N\"urnberg, Sternwartstr. 7, 96049 Bamberg, Germany, wilms@astro.uni-tuebingen.de}

\begin{abstract}

We report on a \chandra~grating observation of the recently discovered hard X-ray transient 
\mbox{IGR~J17497$-$2821}. 
The observation took place about two weeks after the source discovery at a flux level of about 20\,mCrab in the 0.8-8\,keV range. 
We  extracted the most precise X-ray position 
of \mysou, $\alpha_{J2000}$=17$^{h}$ 49$^{m}$ 38$^{s}$.037, 
\mbox{$\delta_{J2000}$= -28$^{\circ}$ 21$^{\prime}$ 17$^{\prime \prime}$.37} 
(90\% uncertainty of 0$^{\prime\prime}$.6). We also report on optical and near infra-red photometric follow-up 
observations based on this position.
With the multi-wavelength information at hand, we discuss 
the possible nature of the source proposing that \mysou~is a low-mass X-ray binary, most likely hosting a black hole,
with a red giant K-type companion.

\end{abstract}



\section{Introduction}
On 2006 September 17 a new hard-X ray transient, \mysou~\citep{soldi06}, 
was discovered by the IBIS telescope \citep{ubertini03} on-board  
\mbox{{\it INTEGRAL}} \citep{winkler03}.
The source was first detected at a flux of about 25 mCrab in the 20--40\,keV range and 
further observations \citep{shaw06,kuulkers06} indicated that \mbox{IGR~J17497$-$2821} was brightening with a 
3--200\,keV \integral~spectrum  well fitted by an absorbed power-law with  $\Gamma$=1.93$\pm$0.05.
Two days later, a \swift~observation was performed 
and a \swift/XRT position at $\alpha_{J2000}$=17$^{h}$ 49$^{m}$ 38$^{s}$.1,
$\delta_{J2000}$= -28$^{\circ}$ 21$^{\prime}$ 16$^{\prime \prime}$.9
with uncertainty of 5${^{\prime \prime}}$.3 radius (90\% containment)
was reported \citep{kennea06}. The XRT spectrum was well fitted using
an absorbed power-law with 
\nh=(4.8 $\pm$ 0.3)$\times$10$^{22}$ cm$^{-2}$ and  $\Gamma=1.6~\pm~0.1$.
The source was also detected by \emph{RXTE}/PCA  
\citep{markwardt06}
and \emph{Suzaku} \citep{itoh06}.
In the quest to understand the nature of this new transient, 
archival optical and follow-up Near Infra-Red (NIR) observations were reported \citep[][respectively]{laycock06,chaty06a,chaty06b}.
Nevertheless, the location of the source near the Galactic center
made it very difficult to assess a unique optical-NIR identification and a sub-arcsec
accuracy of the X-ray source was clearly needed \citep{paizis06b,torres06}.
In this letter we present a \chandra~grating observation of \mysou~together 
with  optical-NIR observations\footnote{The optical and NIR observations were obtained as part of 
the European Southern Observatory (ESO) Target of Opportunity program 078.D-0268 (PI S. Chaty).}
based on the X-ray position (sub-arcsec accuracy) obtained with our \chandra~data.

\section{Observations and data analysis}

\subsection{\chandra~data}

We observed \mysou~for 19\,ksec with \chandra~on 2006 October 1 from 
17:42 UT until 23:42 UT (MJD 54010). 
The High Energy Transmission Grating Spectrometer, HETGS \citep{canizares00}, was used.
It  has two sets of gratings, the High Energy Grating, HEG 0.8--10\,keV, 
and Medium Energy Grating, MEG 0.4--8.0\,keV.
We reduced the data in a standard manner, using the CIAO version 3.3 software 
package and \chandra~CALDB version 3.2.3. The spectra were analyzed with 
the \isis~analysis system, version 1.3.0 \citep{houck02}. 
We obtained the X-ray position of the source using the  \emph{findzero.sl} 
routine\footnote{http://space.mit.edu/CXC/analysis/findzo/index.html},  strongly 
recommended by the \chandra~grating team in case of bright sources for which the zeroth-order (un-dispersed) position is 
piled-up (about 47\% pile-up in the present case).
We extracted 
the first order dispersed spectra ($m=\pm1$ for HEG and MEG, for a total of four spectra) and to 
increase the signal-to-noise ratio, we merged the two HEG ($m=\pm1$) and MEG ($m=\pm1$) spectra into two 
final combined spectra. 
We  binned the data to obtain a minimum of 16 counts per bin for
both HEG and MEG (0.8--8\,keV), as well as a minimum number of 16 channels per 
bin for HEG and 8 for MEG.

\begin{figure*}
\includegraphics[width=1.0\linewidth]{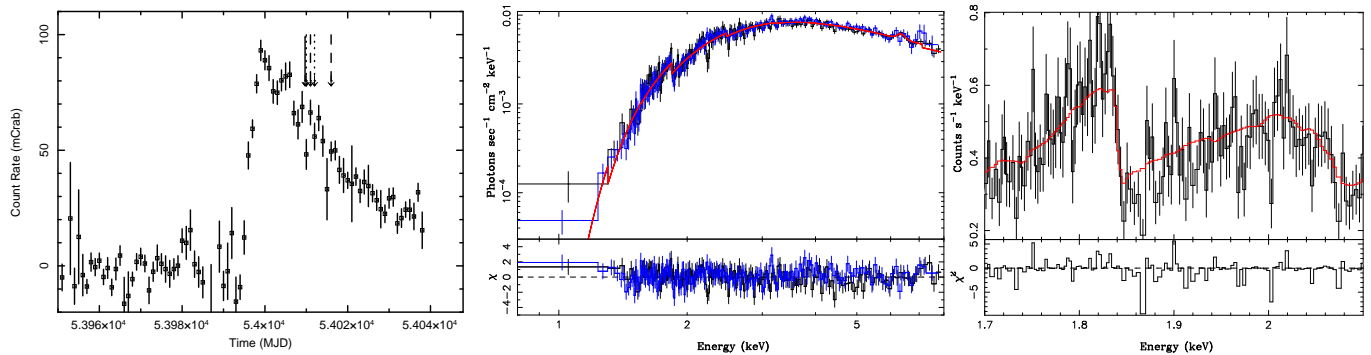}
\caption{\emph{Left panel:}  X-ray intensity history of the 2006 outburst 
of IGR~J17497$-$2821 provided by the {\it \mbox{Swift}}/BAT team (15--50\,keV data). The time of our follow-up observations is shown:
solid line arrow for \chandra, dotted for NIR and dashed for optical observations. \emph{Middle panel:} unfolded 
\chandra~ spectrum with best fit model ({\sc phabs(diskbb+po+gaussian)}, Table~1) superimposed. Residuals between the model and the data in units of  1$\sigma$
are shown.  \emph{Right panel: zoom in the Si~XIII absorption line region (3$\sigma$, MEG and HEG combined)}. \label{fig:spectrum}}
\end{figure*}

\begin{figure*}
\includegraphics[width=0.5\linewidth]{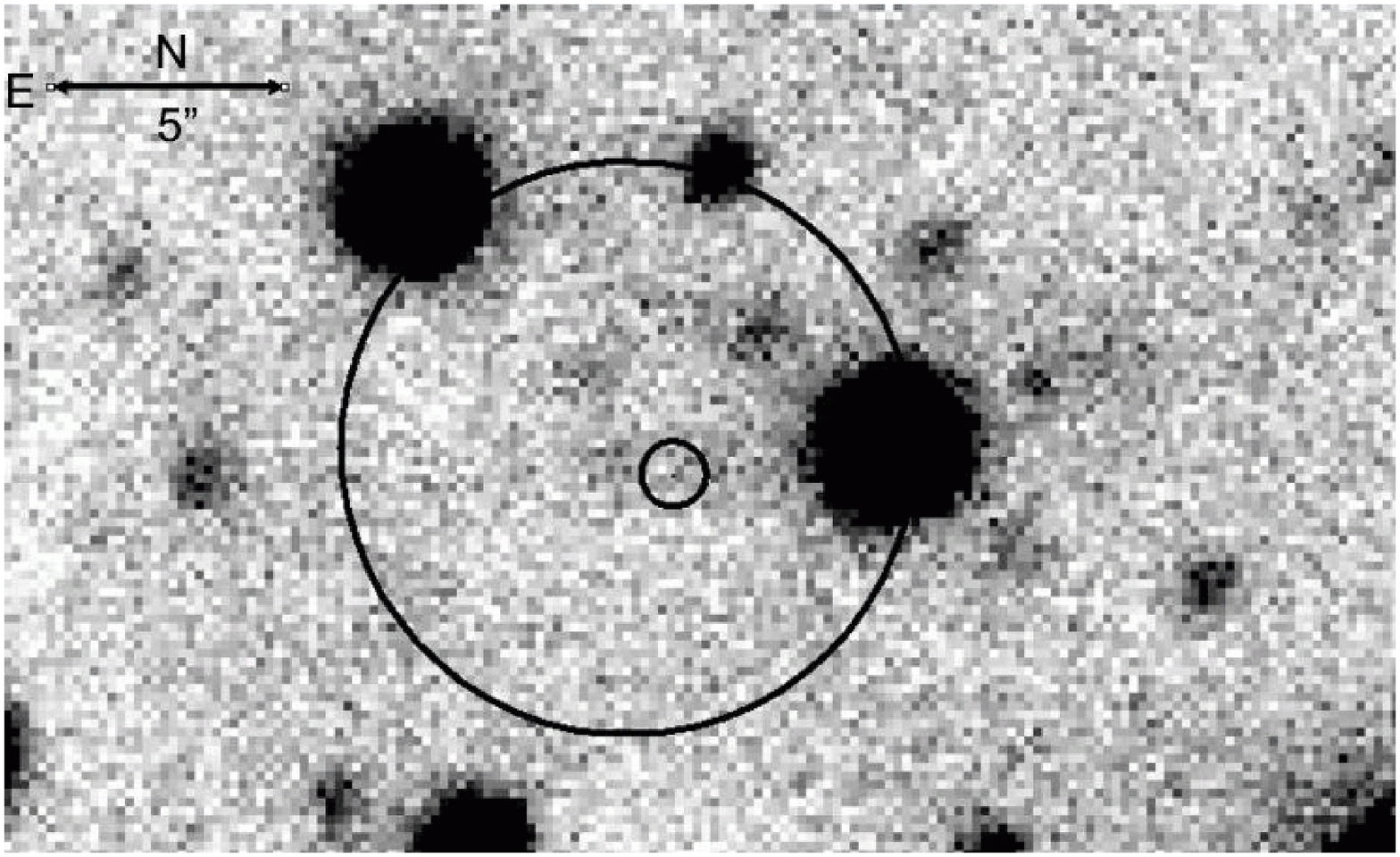}
\includegraphics[width=0.5\linewidth]{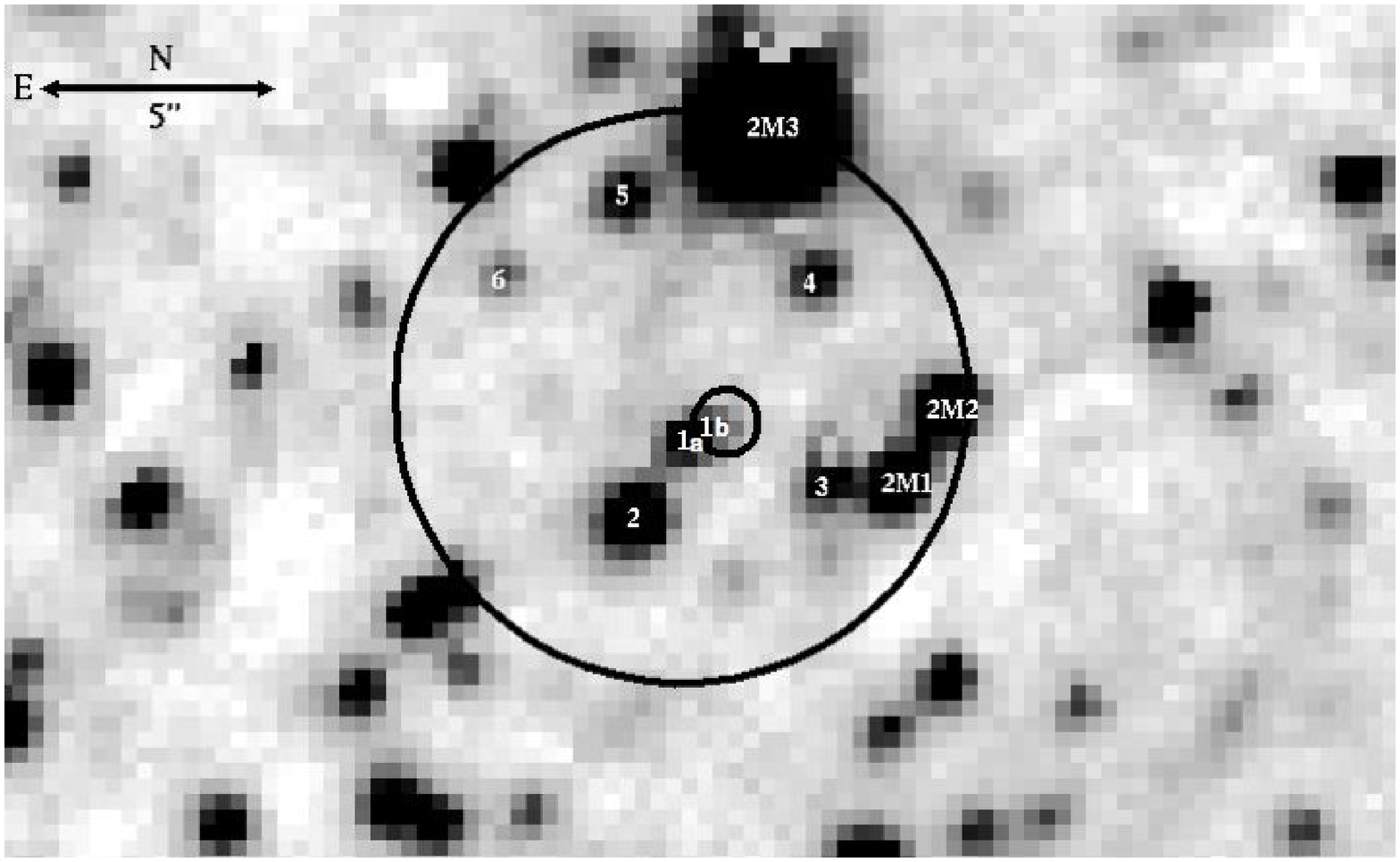}
\caption{\emph{Left panel}: $Z$-band image of the field of view of IGR~J17497$-$2821.
\emph{Right panel}: $K_{s}$-band image of the field of view of IGR~J17497$-$2821.
In both images we over-plot the 
\swift/XRT 90\% error circle, 5$^{\prime \prime}$.3 radius \citep{kennea06}, 
and the \chandra~90\% error circle, 0$^{\prime \prime}$.6 radius (this paper). 
North is up, East is to the left, and the length of the arrow represents 
5$^{\prime \prime}$. \label{fig:coun}}
\end{figure*}

\subsection{Optical and NIR data}
We also obtained optical photometry in $U$ ($357.08$\,nm), $B$ ($421.20$\,nm), 
$V$ ($544.17$\,nm), $R$ ($641.58$\,nm), $I$ ($794.96$\,nm) and $Z$
($840.90$\,nm) bands of the field of view (FoV) of \mbox{IGR~J17497$-$2821} with the
imager SUSI2, installed on the 3.5\,m New Technology Telescope (NTT) of
La Silla observatory (ESO, Chile).  The observations were performed 
on 2006 October 3 between 00:01 UT and 00:12 UT and
2006 October 7 between 23:55 UT and 8 00:06 UT.  We used
the large field imaging of SUSI2's detector with 0$^{\prime \prime}$.166 pixel$^{-1}$ 
image scale and  5$^{\prime}$.5$\times$5$^{\prime}$.5 FoV.  The images
were binned by a factor 2, and the integration time was of 60\,s
for each exposure. 
 Photometry was preformed relative to the standard star Mark-A of \cite{landolt92}.

We also obtained NIR photometry in $J$ ($1.247\,\mu$m), $H$ ($1.653\,\mu$m) 
and $K_{s}$ ($2.162\,\mu$m) bands of the field around \mysou~with the 
spectro-imager SofI, installed on the NTT.  The
observations were performed in two epochs, the first between 2006 October
1, 23:48 UT, and October 2, 00:11 UT, and the second between 
October 3, 23:45 UT, and October 4, 00:09 UT.  We used the large
field imaging of SofI's detector with 0$^{\prime \prime}$.288 pixel$^{-1}$ image scale and 
a 4$^{\prime}$.94$\times$4$^{\prime}$.94 FoV.  
To accurately determine the sky brightness, the observations consisted of nine 
slightly (30$^{\prime \prime}$) dithered frames of 10\,sec each per filter, for a total 
exposure of 90\,sec in each filter. 
 Photometry was preformed relative to the standard stars  sj9181 and sj9183
 of \cite{persson98}.

We used the Image Reduction and Analysis Facility (\rm{IRAF}) suite to
perform data reduction, carrying out standard procedures of optical
and NIR image reduction. Aperture photometry in the crowded field of \mysou~was performed with the
\rm{daophot} package of the \rm{IRAF} suite.  This package
allows one to build a synthetic Point Spread Function (PSF) using bright
and isolated stars in the field of view, and then to scale and
subtract this PSF from all the stars. With an iterative PSF subtraction  
it is possible to reveal fainter objects which are
hidden in the glare of brighter ones, such as blended objects.
\begin{table*}
\tiny
\begin{tabular}{ccccccccc}
\hline
\hline
{\nh} &{$\Gamma$} &  {kT$_{in}$ } &  {R$_{in}$} &{Line\_en} &{Line\_width}&{Line\_eq-width}& {Flux$_{0.8-8\,keV}$}&\chis (d.o.f.)\\
{($\times10^{22}$\,cm$^{-2}$)} &  & {(keV)} & {(km, at 8kpc)} & {(keV)}&{(keV)}& {(eV)} &{(\ergcms)}& \\
\hline
\hline
{4.4$\pm$0.1} & {$1.23\pm0.05$} & - & -& - &- &-& {2.95$\times10^{-10}$}& {1.17 (335)} \\
\hline
{5.6$\pm$0.4} &{1.5$\pm$0.1}  & {0.19$\pm$0.02} & {$280{+118 \atop -114}$/$\sqrt(cos\theta)$} &
{6.4 $\pm$0.1} &  {$<$0.22}& {$45.4{+0.1 \atop -32.1}$}&{2.90$\times10^{-10}$}&{1.05 (330)}\\
\hline
\end{tabular}
\label{fit}
\caption{Summary of the spectral fit parameters (0.8--8\,keV). }
\end{table*}

\begin{table}
\tiny
\begin{centering}
  \begin{tabular}{ccccc}
      \hline
      \hline
Candidate      &  mag$_{J}$ & mag$_{H}$ &mag$_{Ks}$ & mag$_{V}$ \\
\hline
\hline
\emph{1a}  &              20.6$\pm$0.2 &17.5$\pm$0.2 & 14.7$\pm$0.1 & \\
\emph{1b} &            -  & - &  16.1$\pm$0.3 & \\
\hline
Blend \emph{1a}+\emph{1b}     &  -  & - & - & $>$23\\
     \hline
\label{mag}
  \end{tabular}
\end{centering}
\caption{Summary of the optical and NIR results. }
\end{table}

\section{Results}
A hard X-ray intensity history of the 2006 outburst of \mysou~seen by \swift/BAT is presented in 
Fig.~\ref{fig:spectrum}, left panel. The time of our \chandra, 
optical and NIR  observations are shown by the indicated arrows.

\subsection{X-ray position and grating spectrum}
We extracted the X-ray position of \mbox{IGR~J17497$-$2821} from the zeroth-order image
obtaining  $\alpha_{J2000}$=17$^{h}$ 49$^{m}$ 38$^{s}$.037, 
\mbox{$\delta_{J2000}$= -28$^{\circ}$ 21$^{\prime}$ 17$^{\prime \prime}$.37}.
Given the brightness of the source, the statistical error is smaller than the 
absolute position accuracy of \chandra, 0$^{\prime \prime}$.6 at 90\% 
uncertainty\footnote{http://cxc.harvard.edu/cal/ASPECT/celmon/}. Therefore we attribute to the position 
found a 90\% uncertainty  of 0$^{\prime \prime}$.6.
This position, compatible with the \swift/XRT one \citep{kennea06}, 
was immediately announced to the community by \cite{paizis06b}.
The first order HEG and MEG spectra of \mysou~are shown in Fig.~\ref{fig:spectrum}, middle 
panel. 
An absorbed power-law with column density
 \nh=(4.4$\pm$0.1)$\times$$10^{22}$\,cm$^{-2}$ ({\sc phabs} model)
 and photon index $\Gamma=1.23\pm0.05$ fits the data  
with a reduced \chis=1.17 for 335 d.o.f in the 0.8--8\,keV band (Table~1).
The absorbed flux is about 3$\times$$10^{-10}$\ergcms~in the 0.8--8\,keV range ($\sim$20mCrab) with no discrete features evident in the spectrum, aside
from the ISM absorption edges associated with Si and S.
The most significant feature in the \emph{combined residuals} is a hint of
 He-like Si absorption line at 1.867\,keV
(Si~XIII, 3$\sigma$ level, Fig.~\ref{fig:spectrum}, right  
pane), likely associated to the ISM in its hot-diffuse phase. If confirmed, this is the first 
detection of Si in the hot ISM \citep{wang05}.
The power-law slope of $\sim$1.2  is harder 
than the measurements currently available from the other missions  \citep[$\Gamma=1.6$,][]{kennea06,itoh06,rodriguez06b,walter07}.
The source is rather 
constant during our observation and instrumental issues such as a high pile-up 
are very unlikely (pile-up fraction in the HETGS arms 
of about 2\%).
Most likely, the heavy absorption of the system leaves a \chandra~effective
energy range that is very narrow (1.5-8\,keV), and we are systematically
underestimating both the spectral slope and the absorption column density.
The power-law model alone does not seem to be the best fit model for the \chandra~grating spectra 
and adding a thermal 
component (\diskbb) to the power-law 
results in a much better fit (probability of $\sim$10$^{-7}$ that the improvement was purely due to chance).  
Table~1  shows 
the spectral parameters we obtain with the new model to which we added also a marginally detected 
6.4 keV iron line ($\Delta$\chiss=9 for the three additional line parameters).

\subsection{Astrometry}
We computed the astrometry of the $Z$ and $K_{s}$ band images of \mysou, 
respectively taken on 2006 October 8, 00:05 UT, and
2006 October 2, 00:10 UT.  The optical astrometry has been computed
with all USNO stars located in the field, 
while the
NIR  astrometry used all 2MASS stars located in the field. 
The rms error on the astrometry is respectively of 0$^{\prime \prime}$.35
for the optical image and 0$^{\prime \prime}$.32 for the NIR one.
The field of view of \mysou~in the optical ($Z$ band)
and in the NIR ($K_{s}$ band) are shown in Fig.~\ref{fig:coun}, left and right 
panel respectively. 
We over-plot the \swift/XRT 90\% error circle, 5$^{\prime \prime}$.3
radius \citep{kennea06}, and the \chandra~90\% error circle,
0$^{\prime \prime}$.6 radius (this paper).
The inspection of the $Z$ image (Fig.~\ref{fig:coun}, left) shows that there is no visible
counterpart inside the \chandra~90\% error circle. 
In the $K_{s}$ band image (Fig.~\ref{fig:coun}, right), inside the \swift/XRT error circle,
there are three 2MASS sources (2MASS~17493780--2821181, 2MASS~17493774--2821173 and 
2MASS~17493798--2821120, respectively labeled 2M1,
2M2 and 2M3) and some additional sources which have been labeled
according to \cite{chaty06b}.  Source \#1 is in fact a blended source \citep{torres06}.  
We call \emph{1b} the  new source
compatible with our \chandra~position (0$^{\prime \prime}$.1 away) and \emph{1a} the other one of the blend (0$^{\prime \prime}$.612 away).

\subsection{Photometry}
The iterative cleaning process described in Section 2.2 allowed us 
to extract magnitudes even for the blended objects shown in 
Fig.~\ref{fig:coun}, right panel. 
The obtained instrumental magnitudes were transformed into apparent magnitudes
with the standard method described in \cite{massey92}.
There is no object detected inside the \chandra~error circle in the optical
 images up to a magnitude of V$\sim$23. 
Concerning the NIR, we took the images of the first epoch
(2006 October 1--2) since
for the images of the second epoch the conditions were not photometric,
and the objects \emph{1a} and \emph{1b} could not be de-blended.
In the $J$ and $H$ images the candidate \emph{1b} is not visible, therefore in 
Table~2 we give the apparent magnitudes of the Candidate \emph{1a} in $J$
and $H$, and the apparent magnitudes of both candidates \emph{1a} and \emph{1b}
in $K_{s}$.
The $K_{s}$ magnitude of candidate \emph{1b} is consistent with the $K_{s}$ magnitude
of 15.9$\pm$0.2 reported by \cite{torres06} during observations
performed ten days earlier. The source seems to be quite constant in the NIR, similarly to what reported by \cite{chaty06c}.
There are no archival optical-NIR images of the field at the resolution needed  
and \mysou~will be visible again by the NTT at ESO starting on March 2007.
Follow-up observations are important to detect any variations
from candidates \emph{1a} and \emph{1b}, since even if candidate \emph{1b} is the best
candidate based on its proximity to the \chandra~position, 
only NIR variations correlated with the X-ray flux might
establish which is the real \mbox{counterpart}. 

\section{Discussion}
\mysou~is placed in the direction of the 
 Galactic center, (\emph{l},\emph{b})=(0$^{\circ}$.9,$-$0$^{\circ}$.4), and
we observe a column density of about 5$\times$$10^{22}$\,cm$^{-2}$ that is higher than the 
Galactic average value expected in the source direction, $\sim$1.5$\times$$10^{22}$\,cm$^{-2}$ \citep{dickey90}.
This can imply that there is an additional contribution from within the system. Nevertheless,
the value obtained with the radio maps by \cite{dickey90} does not resolve the small scale non-uniformity of \nh~and 
does not include the possible contribution of molecular hydrogen,  
probably underestimating the true value. 
Given the location in the sky and the high interstellar
absorption, the source is most likely at the distance of the Galactic center or beyond.
For our best fit model, assuming a distance of 8\,kpc, we obtain an (un-absorbed) 1--20\,keV source luminosity that is typical of X-ray binaries, 
$\sim$ $10^{37}$\ergs, similarly to what reported by \cite{rodriguez06b}. The nature of the companion, {\it i.e.} Low Mass X-ray Binary (LMXB) versus 
High Mass X-ray Binary (HMXB) and of the compact object, {\it i.e.} Black Hole (BH) versus Neutron Star (NS) is still a matter of 
debate.
The general X-ray properties seem to suggest that the source is a (transient) LMXB and 
the  X-ray spectrum we obtain is compatible with a LMXB in the so-called low-hard state (LHS).

Using the relation between \nh~and interstellar extinction A$_{V}$ \citep{predehl95} 
and the ratio A$_{K}$/A$_{V}$=0.112 \citep{rieke85}, our observed column density range (4.3--6$\times$$10^{22}$\,cm$^{-2}$,
to allow for both fitting models in Table~1)
corresponds to a $K$ band extinction in the range of A$_{K}$=2.7--3.7 magnitudes. This leads to an absolute $K$ magnitude 
range of $-$1.1$>$M$_{K}$$>$$-$2.2 (assumed 
distance of 8\,kpc and observed $K$ magnitude of mag$_{K}$=16.1, see Table~2). This value \citep[see Fig.~1 in][]{chaty02}
suggests a B-type companion  
in the case of a main sequence star, therefore a HMXB, or
a K-type companion in the case of a red giant, 
therefore a LMXB or, more precisely, the fourth known "symbiotic" LMXB \citep[][]{mattana06,
masetti06}. 
We note that in the LMXB frame (favored by the X-ray properties of the source), one 
could envisage a scenario where the companion star is a stellar type K 
\emph{main sequence} star (dimmer than the red giant K-type option) at 8\,kpc that is rendered brighter 
by the accretion disk contribution (NIR emission from reprocessed X-rays). 
To obtain a main sequence K-type star 
from a red giant K-type star at a fixed distance we need a shift of about 6 magnitudes in M$_{K}$ \citep{chaty02}, gap that should 
be covered by the disk emission alone. 
Such an increase has been seen in other sources, due
to the disk contribution, but they were in the high-soft
state and not, as in the present case, in a low-hard state with a cold disc
for which a 3 magnitude increase would be expected.
Conversely, in the absence of  NIR  disk emission, 
the option of a main sequence star located 
much closer than 8\,kpc is also unlikely. In fact, the needed 6 magnitude shift    
would place the source as close as $\sim$0.4\,kpc leading to a  
1--20\,keV source luminosity during our \chandra~observation of about $10^{34}$\ergs, unusually low for 
a LMXB in outburst.
 
The synergy of the NIR and soft X-ray observations presented here seem to suggest that 
\mysou~is a LMXB with a red giant K-type companion. The 
B-type companion in a HMXB system (even if not favored by the X-ray properties) 
or the LMXB main sequence possibility (with proper interplay of distance and accretion disk contribution) 
cannot be ruled out since the source is too faint for a spectral analysis that would lead to much stronger 
constraints than the magnitude estimates presented here.

Regarding the nature of the compact object we note that up to now no pulsations or type-I X-ray bursts,
that would point to presence of NS in the system,   
have been detected \citep{markwardt06,rodriguez06b}.
Our results  
are consistent with a cold (0.2\,keV) disk 
around a BH 
of \eg~10 solar masses at a distance of 8\,kpc and also the power-law slope 
($\Gamma\sim$1.5) 
is typical of a BH in the low-hard state \citep{belloni04}. However, based on our \chandra~grating spectrum alone, 
we cannot rule out the NS LMXB in the LHS scenario, even if this seems rather unlikely given that \mysou~has been detected up to $\sim$300\,keV
\citep{itoh06}, rather hard for a NS LMXBs. Indeed, also \cite{rodriguez06b} and \cite{walter07} using different
diagnostics than ours, favor a  BH as the primary. We note that \cite{walter07} propose a counterpart that 
lies outside our \chandra~error box (0$^{\prime \prime}$.98 away). Although this identification seems unlikely, 
further investigation is needed to locate the true counterpart. Our identification, if confirmed,
makes \mysou~the first symbiotic (K-type \emph{giant} companion) LMXB hosting a BH.

\acknowledgments

AP thanks S. Soldi, N. Mowlavi, S. Shaw and E. Kuulkers for updates on the {\it INTEGRAL} monitoring of the source.
AP and MN thank Tom Aldcroft for useful discussion on the \chandra~absolute astrometry. 
AP, MDS and PU akcnowledge the Italian Space Agency financial and programmatic support via contract I/023/05/0.

\end{document}